\documentclass[12pt]{article}
\usepackage{epsf}

\newcommand{\beq}{\begin{equation}}
\newcommand{\eeq}{\end{equation}}
\newcommand{\beqcol}{\begin{array}{rcl}}
\newcommand{\eeqcol}{\end{array}}

\begin{document}

{\thispagestyle{empty}

\begin{flushright}  KCL-MTH-97-31 \\ May 1997 \end{flushright}
\vfill
\begin{center}
{\Large 
Polynomial Recursion Equations in Form Factors \\
of ADE-Toda Field Theories}
\\
\vspace{2cm}
{\sc Mathias Pillin}             \\
\vspace{1cm}
Department of Mathematics             \\ 
King's College                     \\
Strand                            \\
London WC2R 2LS, U.K.               \\
\                               \\
e-mail: map@mth.kcl.ac.uk 
\medskip

\vfill
{\bf Abstract}
\end{center}
\begin{quote}
It is shown that the problem of calculating form factors in 
ADE affine Toda field theories can be reduced to the 
nonperturbative recursive calculation of polynomials symmetric 
in each sort of variables. We determine these recursion equations 
explicitly for the ADE series and characterize the polynomial 
solutions by an interplay between the weight space of the 
underlying Lie algebra and representations of the symmetric 
group.
\end{quote}

\eject
}

\setcounter{page}{1}


{\bf 1.} 

\bigskip

The knowlegde of form factors, i.e. matrix elements 
of a local operator in a quantum field theory, gives a deep 
insight into the quantum structure of a given Lagrangian field 
theory. Once one has calculated all form factors it is 
possible to classify the operator content of the quantum 
theory or to calculate correlation functions. 

In the class two-dimensional field theories with factorizable 
scattering it is known \cite{SMIR} that form factors do 
satisfy a system of four axioms which makes the problem 
of calculating them somehow feasible.

Form factors have been studied for several diagonal 
scattering theories in recent years, see \cite{CM, ZAM, 
FMS, DM, KOUB, KM, DM2} and references therein. 
In this paper we shall be interested in affine 
Toda field theories which is a class of two-dimensional 
models with factorizable diagonal scattering for which the 
classical S-matrices are known \cite{BCDS,DOR}. 

Form factors for the $A_1^{(1)}$ and for the $A_2^{(2)}$ 
Toda field theories have been studied in \cite{FMS,KM} and 
\cite{ZAM} respectively. The theories are simple in the 
sense that only one type of particle is present and that 
the S-matrix contains only poles of first order. In fact 
only these first order poles are covered by the axioms 
\cite{SMIR} mentioned above. Recently some form factors 
have been calulated for $A_n^{(1)}$ theories in \cite{OOTA} and 
$D_{2n}^{(1)}$ in \cite{MAP}.

\medskip

In this paper we show that the axioms of \cite{SMIR} can be 
applied with slight modifications to situations with 
many different kinds of particles and S-matrices with 
higher order poles. We prove for the case of ADE Toda field 
theories that, once the minimal form factors are known, 
the only thing to be done is to solve a polynomial recursion 
equation. These equations are exact and do not rely on any 
perturbative treatment! Even though it might be difficult 
to find the solutions explicitly we succeed in characterizing 
them by means of the representation theory of the 
symmetric group.

\bigskip


{\bf 2.} 

\bigskip

Let us define the Lagrangian of affine Toda field theory 
in the following way (cf. \cite{CP,FREE}). 

Let ${\bf g}$ be a rank r Lie algebra of type A, D, or E 
\cite{BOURB}. The set of simple roots of ${\bf g}$ is given by 
$\{ \alpha_i \}$ with $ i \in I = \{ 1,2, \ldots , r \}$ 
and we set $\alpha_0 $ to be the negative of the 
highest root. Let $\hat{I}= I \bigcup \{ 0 \}$. 

The fundamental weights $\lambda_i$ are defined using 
the standard euclidean form $ ( .\; , . )$ on 
${\bf g}$ by the condition $ (\lambda_i, \alpha_j ) = 
\delta_{ij}$. Notice in the definition that we restrict 
our discussion to ADE Lie algebras and therefore we do 
not have to care about long and short roots. 

The Cartan matrix of ${\bf g}$ is given by $C_{ij} = 
( \alpha_i , \alpha_j )$. Because we will make use 
of it below we mention that the inverse of $C$ is 
given by $C^{-1}_{ij} = (\lambda_i , \lambda_j )$.

Define 
integers $k_i$ with $i \in \hat{I}$ by the condition 
$k_0=1$ and $\sum_{i \in \hat{I}} k_i \alpha_i = 0$. 
It is of course known that the Coxeter number is given 
by $h=\sum_{i \in \hat{I}}k_i $. We then set 
$X^{\pm} = \sum_{i \in \hat{I}} \sqrt{k_i} 
X_i^{\pm}$, where the $X_i^{\pm}$ are the generators 
of ${\bf g}$ corresponding to the simple root 
$\alpha_i$. $X^{+}$ and $X^{-}$ commute and are elements 
of the Cartan subalgebra of ${\bf g}$. 

Let $ {\bf {\Phi}} $ be a scalar field taking values 
in the Cartan subalgebra of ${\bf g}$. The Lagrangian 
of the affine Toda field theories is the given by 
the following expression \cite{FREE,CP} which is 
defined in a two-dimensional Minkowski space.

\beq 
{\cal{L}} = {\displaystyle{1\over {2}}} (\partial_{\mu} 
 {\bf {\Phi}}, \partial^{\mu} {\bf {\Phi}}) 
- {\displaystyle{ m^2 \over {\beta}}} 
( e^{ {\rm{ad}}( {\bf {\Phi}} )} X^{+}, X^{-} ),
\label{lagrangian}
\eeq

where ad denotes the adjoint action. The parameters $m$ 
and $\beta$, which we take to be real 
throughout this paper, are the mass scale and the 
coupling respectively. It turns out that for the 
purposes of our study the effective coupling \cite{BS} 

\beq
B(\beta ) = {1\over {2 \pi} } { \beta^2 \over { 1+ \beta^2/4\pi }}
\label{B-def}
\eeq

will be of fundamental importance. 

\bigskip


{\bf 3.}

\bigskip

As implicitly stated above, the fields or particles in an affine Toda field 
theory are in one-to-one correspondence to the nodes in the Dynkin diagram 
of the Lie algebra ${\bf g}$. More precisely \cite{DOR}, a particle 
of species $a$ in a given theory might be identified with the 
fundamental weight $\lambda_a$, for $a \in I$. 

According to the symmetries of the Dynkin diagram, particle 
$a$ has a charge conjugate partner $\bar{a}$ which might 
be identical with $a$. We remark that the theories 
coming from ${\bf g} \in \{ D_{2n}, E_7, E_8 \}$ contain 
only self-conjugate particles. For details see \cite{BCDS}. 

Using this correspondence it can be shown that fusing, i.e. 
non-vanishing three point couplings, of particles $a$, $b$, 
and $c$ occurs if there exist integers $r_1$, $r_2$, and 
$r_3$ such that the following equation holds.

\beq
 w^{r_1} \lambda_a +w^{r_2} \lambda_b + w^{r_3} \lambda_c =0 .
\label{dorey-rule}
\eeq

$w$ denotes the Coxeter element of the Weyl group of ${\bf g}$. 
This equation is known as Dorey's rule \cite{DOR}, see also 
\cite{CP}.

\medskip

If we denote the rapidities of a particle by $\theta$ 
the fusing $a + b \to \bar{c}$ occurs at $ i \theta^{c}_{ab}$. 
We will refer to $ \theta^{c}_{ab}$ as well as to $ u ^{c}_{ab} 
= \theta^{c}_{ab} h/ \pi$ and $\bar{u} = h- u$ as fusing 
angles for the particular fusing. Explicit fomulae for 
these angles can be found in \cite{BCDS,DOR}. It can be shown 
that there exists a correspondence between these angles and 
the integers occuring in (\ref{dorey-rule}). 

\medskip

The classical S-matrices for ADE affine Toda field theories have 
been constructed as a solution of the bootstrap equation 

\beq
S_{d\bar{c}}( \theta ) = S_{da} ( \theta - i \bar{\theta}^{b}_{ac} )
 S_{db} ( \theta + i \bar{\theta}^{a}_{bc} ) .
\label{S-bootstrap}
\eeq

It was shown in \cite{DOR} that these S-matrices can 
be nicely written in the following way.

\beq
S_{ab}( \theta ) = \prod\limits_{p=0}^{h-1} 
( \langle 2p + 1 + \varepsilon_{ab} \rangle_{+(\theta)} 
   )^{( \lambda_a, w^{-p} \phi_b )} .
\label{Smat1}
\eeq

We have been using the following notation. For $i \in I$ we 
define $\phi_i = ( 1- w^{-1} ) \lambda_i $. (These are just 
linear combinations of fundamental root vectors with the 
property that $w\phi_i$ is a negative root \cite{BOURB}). 
The functions $ \langle r \rangle_{+(\theta)} $ are defined 
by

\beq
\langle r \rangle_{+(\theta)} = {\displaystyle{ 
  { { (r-1)_{+(\theta)} (r+1)_{+(\theta)}} \over {  (r-1+B)_{+(\theta)}  
     (r-1+B)_{+(\theta)}  }  }}}, \qquad 
  (r)_{+(\theta)}= {\rm sinh}( (\theta + {i \pi \over{h}} r )/2).
\label{brack-def}
\eeq

Using the standard \cite{DOR,CP} two-coloring of the nodes of 
the Dynkin diagram we set $c(\lambda_a )= 1$ if the node is of 
color white and $c(\lambda_a )= -1$ if the color is black. 
With this convention we set $\varepsilon_{ab} 
= ( c(\lambda_a) - c( \lambda_b) )/2 $. 

For later purposes the following form of the S-matrix, completely 
equivalent to (\ref{Smat1}), is useful.

\beq
S_{ab}(\theta ) = {\displaystyle{ {\xi_{ab}(- \theta )\over{ 
 \xi_{ab}( \theta )}}}}, \qquad 
  \xi_{ab}(\theta ) = \prod\limits_{p=0}^{h-1} \langle p 
    \rangle_{+(\theta )}^{m_{ab}(p)} .
\label{Smat2}
\eeq

Comparing this form with (\ref{Smat1}) it is clear that the 
integers $m_{ab} \ge 0 $ can be easily related to the exponent 
in (\ref{Smat1}). We list a few propeties of this quantity.

\beq
\beqcol
m_{ab}(1) &=& \delta_{ab}, \\ 
m_{ab}(h-p) &=& m_{\bar{a}b}(p)
\eeqcol  
\qquad
\beqcol
m_{ab}(-p) &=& - m_{ab}(p), \\
m_{\bar{c}d}(p) & = & 
m_{ad}(p-\bar{u}^{b}_{ac} ) +  m_{bd}(p+\bar{u}^{a}_{bc} ).
\eeqcol
\label{m-ident}
\eeq

The last equation is of course a descendant of the bootstrap 
equation (\ref{S-bootstrap}). 

\bigskip


{\bf 4.}
 
\bigskip

The object of study in this paper are the form factors in 
an ADE Toda field theory. This is a matrix element of a 
local operator ${\cal{O}}(x)$. Without loss of generality we can 
take the local operator at the origin of our space-time and 
allow the matrix element to be taken for several incoming 
particles and the vacuum.

\beq
F_{a_1 \ldots a_n}(\theta_1,\ldots , \theta_n) := 
 \langle 0 | {\cal{O}}(0) | \theta_1,\ldots , \theta_n \rangle .
\label{formdef}
\eeq

In our conventions the indices $a_i$ denote the species 
of a particle of rapidity $\theta_i$.

It is know that the form factor in the class of theories we 
are considering is subject to four axioms 
\cite{SMIR}. The first two of them are known as Watson's 
equations. One of them reflects the fact that particles 
$a_i$ and $a_{i+1}$ can be interchanged by the S-matrix 
$S_{a_{i}a_{i+1}}(\theta_i - \theta_{i+1})$ while the other 
one is just a monodromy property of $F$. 

The solutions of Watson's equations can be constructed by 
the following procedure. Since $F$ is a meromorphic function 
in several variables we can split it locally into a part, which 
we will denote by $K$, containing poles and zeros in a given 
region, times another 
part which is analytic without zeroes there. We perform 
this split in the strip $ 0 < {\rm Im}\; \theta < 2 \pi $.

\beq
F_{{a_1 \ldots a_n}}  ( \theta_ 1, \ldots \theta_ n) = 
K_{{a_1 \ldots a_n}}  ( \theta_ 1, \ldots \theta_ n) \prod_{i<j} 
F^{{\rm min}}_{a_i a_j} ( \theta_ i- \theta_ j ) ,
\label{split}
\eeq

$F^{{\rm min}}_{a_i a_j} ( \theta_ i- \theta_ j )$ is known as 
the minimal form factor. It can be shown that once one assumes 
some obvious monodromy properties for $K$ (mainly periodicity), 
this object drops out of Watson's equations and 
$F^{{\rm min}}_{a_i a_j} ( \theta_ i- \theta_ j )$ then satisfies 
a simple two-particle equation. The minimal solutions in the 
ADE-case are known 
and can be given either in an integral form \cite{MAX} 
or in a product expansion of $\Gamma$-functions \cite{OOTA}.

\smallskip

Using these solutions one can show that the two remaining 
axioms lead to equations for $K$ only. One of these equations 
arises from the fact that we have 
particles and their conjugate partners in our theory, and 
reflects a zero angle scattering of a given particle 
with all the other fields approaching the local operator, 
for details see the appendix of \cite{SMIR}. We then obtain 
the following kinematical residue equation.

\beq
\beqcol
& & -i \ {\rm res}_{\theta^{\prime}=\theta+i\pi} \ 
K_{\bar{a}ad_{1} \ldots d_{n}}
(\theta^{\prime}, \theta,\theta_ {1}, \ldots, \theta_ {n})   =               \\
&=& K_{d_{1}\ldots d_{n}}(\theta_ {1}, \cdots, \theta_ {n})
\left(\prod_{j=1}^{n} \xi_{ad_{j}}(\theta-\theta_ {j})-
\prod_{j=1}^{n} \xi_{ad_{j}}(\theta_ {j}-\theta) \right) /
F^{{\rm{min}}}_{\bar{a}a}(i\pi).
\eeqcol
\label{K-kinemat}
\eeq

The last condition to be imposed is a consequence of the 
presence of intermediate bound states, i.e. fusings. In 
\cite{SMIR} this condition was established only for 
fusings arising from first order poles of the S-matrix. We 
will show below that with some minor modifications this 
last axiom remains valid in the presence of any odd order 
forward channel pole of the S-matrix. For a related statement 
see \cite{DM2}. The bound state residue equation for $K$ is then: 

\beq
\beqcol
& & -i \ {\rm res}_{\theta^{\prime} = \theta+ i \theta_{ab}^c} \ 
K_{abd_{1}\ldots d_{n}} (\theta^{\prime}, \theta, \theta_ 1, \cdots,
\theta_ n)               =             \\
&=& \Gamma_{ab}^c K_{\bar{c} d_1 \ldots d_n}
(\theta+i\bar{\theta}_{bc}^a, \theta_ 1, \cdots, \theta_ n)
\prod_{j=1}^n  \lambda_{ab;d_j}^c (\theta+i\bar{\theta}_{bc}^a-\theta_ j)
/ F^{{\rm{min}}}_{ab}(i\theta_{ab}^c).
\eeqcol
\label{K-bound}
\eeq

Here $\Gamma_{ab}^c= - i {\rm{res}}_{\theta = i \theta^{c}_{ab}} 
S_{ab}(\theta )$ is the vertex. The object 

\beq
\lambda^c_{ab;d}(\theta)^{-1} = 
{ F^{{\rm min}}_{ad} (\theta+ i \bar{\theta}^b_{ac} ) 
  F^{{\rm min}}_{bd}(\theta- i \bar{\theta}^a_{bc} ) \over 
{F^{{\rm min}}_{\bar{c} d} (\theta) }} = 
\prod_{p=0}^{\bar{u}^b_{ac}} 
      \langle \bar{u}^b_{ac}- p \rangle_{+(\theta)}^{m_{ad}(p)} 
\prod_{p=0}^{\bar{u}^a_{bc}-1} 
      \langle p -\bar{u}^a_{bc} \rangle_{+(\theta)}^{m_{bd}(p)} ,
\label{lamdef}
\eeq

is directly related to (\ref{S-bootstrap}) and will 
be of importance later in the paper. For that pupose 
we note the following identity \cite{OOTA}.

\beq
\lambda^{c}_{ab;\bar{d}}(\theta - i \theta^{c}_{ab} )
\lambda^{c}_{ab;d}(\theta - i \bar{\theta}^{b}_{ac} )
= \xi_{ad}(- \theta) . 
\label{lam-ident}
\eeq

\bigskip


{\bf 5.}

\bigskip

Based on the results in \cite{OOTA,MAP} we will now propose 
a general ansatz for the $K$-part of the form factor. We denote 
by $N_k$, $k \in I$, the number of particles of 
type $k$ in the expression (\ref{formdef}). Rather than to 
work with the rapidities $\theta_i$ itself it is possible 
and useful to take $x_i= e^{\theta_i}$ as basic variables. 
All these $x_i$'s of a given species will be grouped together 
in the vector

\beq 
{\bf x}^{(k)} = ( x_1^{(k)}, x_2^{(k)}, \ldots , x_{N_k}^{(k)} ), 
\qquad k \in I .
\label{vecdef}
\eeq

The singularities leading to the kinematical residue are 
parametrized by expressions of the form $(x_i + x_j)^{-1}$ 
while the ones corresponding to fusings are contained in 
functions of the form \cite{OOTA}

\beq
W_{k l}(x^{(k)}_i , x^{(l)}_j ) =  \prod\limits_{p=0}^{h-2}
( x^{(k)}_i - \Omega^{p+1} x^{(l)}_j )^{m_{kl}(p)} \ 
( x^{(k)}_i - \Omega^{-p-1} x^{(l)}_j )^{m_{kl}(p)} , 
\qquad \Omega= e^{i \pi /h}.
\label{W-def}
\eeq         

Notice that the product only goes up to $h-2$ to ensure 
consistency of this ansatz \cite{OOTA}.

With this notation the parametrization of $K$ is the following.

\beq
\beqcol
K_{[N_1, \ldots , N_r ]} ( {\bf{x}}^{(1)}, \ldots, {\bf{x}}^{(r)} ) 
&=& Q_{[N_1, \ldots , N_r ]} ( {\bf{x}}^{(1)}, \ldots, {\bf{x}}^{(r)} )
 \    \prod\limits_{k\in J} \prod\limits_{i,j} 
    { 1 \over{x_i^{(k)} + x_j^{(\bar{k})} }} \times    \\
& & {\hspace{-2cm}} \times  \prod\limits_{k=1}^{r} \prod\limits_{i<j}^{N_k} 
    { 1 \over {W_{kk}(x_i^{(k)},x_j^{(k)} )}}
\times \prod\limits_{k=1}^{r-1} \prod\limits_{l=k+1}^{r} 
       \prod\limits_{i=1}^{N_k}  \prod\limits_{j=1}^{N_l} 
     { 1 \over {W_{kl}(x_i^{(k)},x_j^{(l)} )}} .
\eeqcol
\label{K-def}
\eeq 
 
For the kinematical pole part we have been introducing the index set 
$J$ which contains the indices of all self-conjugate particles, and 
the indices of the non-self-conjugate ones, such that only 
the particles and not their conjugates are being counted.

\medskip

The ansatz (\ref{K-def}) is chosen in such a way that it contains 
in an explicit way all informations about the poles necessary for 
the residue equations (\ref{K-kinemat}) and (\ref{K-bound}). In 
other words we have reduced the problem of calculating the form 
factor to the problem of calculating the $Q$'s in (\ref{K-def}).

We are going to prove that 
$Q_{[N_1, \ldots , N_r ]} ( {\bf{x}}^{(1)}, \ldots, {\bf{x}}^{(r)} )$ 
is a polynomial, which is symmetric in at least all the components 
of the vectors ${\bf{x}}^{(k)}$. 

\medskip

Let $A_{ab}$ be the set of integers $p$ in the expression 
for the S-matrix (\ref{Smat2}) which have $m_{ab}(p) > 0$, such 
that $p$ occurs $m_{ab}(p)$ times in $A_{ab}$. We denote 
the number of elements in $A_{ab}$ by $ \# A_{ab}$. Since 
this number will be frequently used in the sequel we will 
show the following.

\bigskip

{\bf Proposition 1:} {\it Let $C_{kl}^{-1}$ as in section 2 denote 
an entry in the inverse of the Cartan matrix of the Lie algebra 
${\bf{g}}$. We then have:

\beq 
  \# A_{ab} = C_{ab}^{-1} + C_{a\bar{b}}^{-1} .
\label{prop1}
\eeq

}

\medskip

{\sc Proof:} Considering the exponent in (\ref{Smat1}) we 
can see that $\sum_{p=0}^{h-1} (\lambda_a, w^{-p} \phi_b) 
= \sum_{p=0}^{h-1} (\lambda_a, (w^{-p}-w^{-p-1}) \lambda_b ) 
= 0 $ because of the cyclicity of the Coxeter element. 
Having a closer look at (\ref{Smat1}) and (\ref{Smat2}) one 
can see that for a any $p \in \{ 1,2, \ldots , h-1 \} $ 
we have with an entry $ \langle p \rangle $ always an 
entry of the form $ \langle 2h -p \rangle^{-1} = 
\langle -p \rangle^{-1}$. These two facts show that 
in order to compute $ \# A_{ab}$ we only have to consider 
the ``positive'' $\langle p \rangle $-entries, hence to evaluate the sum

\beq
 \# A_{ab} = \sum_{p=0}^{s} (\lambda_a, (w^{-p}  - w^{-p-1} ) 
\lambda_b ) =  (\lambda_a, \lambda_b) - (\lambda_a, w^{-s-1} \lambda_b ).
\label{A-anz}
\eeq

Due to (\ref{Smat1}) $s$ is given by $ s = h/2 -1 $ for ADE Lie 
algebras apart from $A_{2n}$ while for $A_{2n}$ which has an 
odd Coxeter number we have $s = h/2 -1 -\varepsilon_{ab}$. 

In the Weyl group of a Lie algebra there is a longest element $w_0$ 
with the property $w_0 \lambda_a = - \lambda_{\bar{a}}$, for 
$a \in I$. This special element can be related to the Coxeter 
element in a standard way \cite{CP}. We are using this result in 
a weak sense which leads in all cases to 

\beq
(\lambda_a, w^{-s-1} \lambda_{b} ) = (\lambda_a , w_0 \lambda_b ) 
= - ( \lambda_a , \lambda_{\bar{b}} ) . 
\eeq

Using this relation in (\ref{A-anz}) together with the fact 
mentioned in section 2 that the inner product of two fundamental 
weights is just an entry in the inverse of the Cartan matrix 
establishes the Proposition. 

\bigskip

We are now in a position to show the main result of this section, 
which is, of course, strongly supported by the results of 
\cite{KM,ZAM,OOTA,MAP}.

\medskip

{\bf Theorem 1:} {\it The object $Q_{[N_1, \ldots N_r ]}( {\bf{x}}^{(1)} , 
\ldots , {\bf{x}}^{(r)} ) $ appearing in the ansatz (\ref{K-def}) 
is a polynomial for all ADE affine Toda field theories. 
}
\medskip

{\sc Proof:} In order to verify the Theorem we have to show that 
by inserting the ansatz for $K$ in (\ref{K-def}) into the 
kinematical (\ref{K-kinemat}) and the bound state (\ref{K-bound}) 
residue equations we get equations for the $Q$'s which are polynomial. 

The first observation to make is that due to the 
structure of the polynomial in the denominator of $K$ (\ref{K-def}) 
it is possible 
to seperate there the variables involved in the limiting process 
for the residues in (\ref{K-kinemat}) and (\ref{K-bound}) 
from those being only spectators in the 
process. This means that by this procedure we split the 
denominator polynomial in all $N+2$ variables 
into and a polynomial in $N$ variables not containing the 
variables involved in taking the residue and another polynomial 
in $N+2$ variables. 

The actual proof of our result basically consists of a 
tedious calculation which makes several times use of the 
identities (\ref{m-ident}) and the explicit form of the 
S-matrix (\ref{Smat1}) and (\ref{Smat2}). 

In order to write down the result of the calculation as 
transparent as possible we again have to introduce some 
notation. The residues in (\ref{K-kinemat}) and (\ref{K-bound}) 
are taken in the vicinity of the rapidity $\theta $ which 
in our exponential language of (\ref{vecdef}) corresponds 
to the variable $x= e^{\theta}$. Evaluating the 
residue for $W_{kl}$ in (\ref{W-def}) suggests to 
introduce the symbol

\beq
  [ p ]^{(k)}_i = x - \Omega^p x^{(k)}_i, \qquad
  k \in  I , \qquad i \in \{ 1, \ldots , N_k \} .
\label{brack-def-1}
\eeq

From this symbol we can construct the following compound 
in a form in which it will appear in the equations for the $Q$'s.  

\beq
(\{ p \}^{(k)}_i)^{\mu_{ab}(p)} := 
 ( [ h-p -1]^{(k)}_i  [ h-p + 1]^{(k)}_i )^{m_{ab}(h-p)} 
 ( [ -p -1+ B ]^{(k)}_i  [ -p + 1-B]^{(k)}_i )^{m_{ab}(p)}. 
\label{brack-def-2}
\eeq

$B$ is the effective coupling of the Toda field theory as defined
in (\ref{B-def}). From the kinematical equation (\ref{K-kinemat}) we derive:

\beq
\beqcol
Q_{\bar{a} a [N_1, \ldots , N_r ]} ( -x, x; {\bf{x}}^{(1)}, \ldots , 
{\bf{x}}^{(r)} ) &=& -i (-1 )^{N_a} \ x^{2 (C_{\bar{a}a}^{-1} + 
C_{aa}^{-1}) -1 } \ 
{ \prod_{p=1}^{h-2} ( 2 {\rm cos}({{p+1}\over{2h}} \pi ) 
)^{2m_{\bar{a}a}(p)} \over 
{F^{\rm{min}}_{\bar{a}a}( i \pi )} }   \times                            \\
      & &                                                          \\
\prod\limits_{i=1}^{N_a} {\displaystyle{ {1}\over{ [0]^{(a)}_i }}} 
                 \times                                            
\prod\limits_{i=1}^{N_{\bar{a}}} 
{\displaystyle{ {1}\over{ [h]^{(\bar{a})}_i }}} 
                 & \times &
\left( \prod\limits_{k=1}^{r} \prod\limits_{i=1}^{N_k} 
       \prod\limits_{p=1}^{h-1} ( \{ p \}^{(k)}_i )^{\mu_{ab}(p)}
                   -
       \prod\limits_{k=1}^{r} \prod\limits_{i=1}^{N_k} 
       \prod\limits_{p=1}^{h-1} ( \{ -p \}^{(k)}_i )^{\mu_{ab}(p)}
\right) \times                                                      \\
                  & &                                               \\
 & & {\hspace{-2cm}}   \times 
 Q_{ [N_1, \ldots , N_r ]} ( {\bf{x}}^{(1)}, \ldots , {\bf{x}}^{(r)} ).
\eeqcol
\label{kinemat}
\eeq

This is a polynomial recursion equation. The factors $ 1/[0]^{(a)} $ and 
$ 1/[h]^{(\bar{a})}$ are cancelled by corresponding expressions 
in the bracket which contains the difference.  

It remains to state the equation for $Q$ arising from the 
bound state residue equation (\ref{K-bound}). 

\beq         
\beqcol      
Q_{ab [N_1, \ldots N_r]} 
      (x \Omega^{\bar{u}^{b}_{ac}}, x \Omega^{-\bar{u}^{a}_{bc}}; 
                {\bf{x}}^{(1)}, \ldots {\bf{x}}^{(r)}) &=& 
i \; x^{2(C^{-1}_{ab} + C^{-1}_{a\bar{b}})-\delta_{\bar{a} b}} \;
H_{abc} \; \displaystyle{ 
  { { \Gamma^{c}_{ab}} \over { F^{\rm{min}}_{ab}(i \theta^c_{ab} )}}}
\prod\limits_{i=1}^{N_{\bar{a}}} {\displaystyle{ 
  {1}\over{ [u^b_{ac}]^{({\bar{a}})}_i }}} 
                 \times                                      \\
& & {\hspace{-8cm}} 
\prod\limits_{i=1}^{N_{\bar{b}}} {\displaystyle{ 
  {1}\over{ [-u^{a}_{bc}]^{({\bar{b}})}_i}}}  
                 \times         
\prod\limits_{i=1}^{N_c} [ h ]^{(c)}_i 
       \times            
\prod\limits_{k=1}^{r}  \prod\limits_{i=1}^{N_k}
    \left( \prod\limits_{p = 1 }^{\bar{u}^{b}_{ac}} 
          ( \{ \bar{u}^{b}_{ac} - p  \}^{(k)}_i)^{\mu_{ak}(p)}  
 \times \prod\limits_{p =1 }^{\bar{u}^{a}_{bc}-1} 
        ( \{ p - \bar{u}^{a}_{bc} \}^{(k)}_i)^{\mu_{bk}(p)} \right) 
 \times \\
\times Q_{\bar{c}[N_1, \ldots N_r]} 
      (x ; {\bf{x}}^{(1)}, \ldots {\bf{x}}^{(r)})  
\eeqcol
\label{bound}
\eeq

To be complete we give the explicit form of the complicated factor:

\beq
\beqcol
 H_{abc} & = & 
\Omega^{\bar{u}^{b}_{ac} ( N_{\bar{a}} + 2 N + m_{ab}(u^{c}_{ab}-1) )}
\; 
\Omega^{-\bar{u}^{a}_{bc} (N_{\bar{b}} + 2 N )}
\;
(\Omega^{\bar{u}^{b}_{ac} } + 
             \Omega^{-\bar{u}^{a}_{bc}})^{\delta_{\bar{a}b}}  \times  \\
& \times &
 \prod\limits_{p=1}^{h-2} ( \Omega^{\bar{u}^{b}_{ac}} - 
                   \Omega^{- \bar{u}^{a}_{bc}-p-1})^{m_{ab}(p)} \times 
 
 \prod\limits_{ {p=1}\atop{p \ne u^{c}_{ab} -1}}^{h-2} ( \Omega^{\bar{u}^{b}_{ac}} - 
                   \Omega^{- \bar{u}^{a}_{bc}+p+1})^{m_{ab}(p)} 
\eeqcol
\eeq

where $N=\sum_{i=1}^{r} N_i$. 

We have thus found that even the 
bound state recursion equation 
(\ref{K-bound}) leads to a polynomial equation for $Q$. 
The negative powers in (\ref{bound}) are actually cancelled, as 
can be easily seen. 

As a result we have established the theorem.

\bigskip

{\it Remark:} It is known that the S-matrices of ADE Toda field 
theories might have poles which 
correspond to forward channel fusings \cite{BCDS,DOR} in 
order greater than one. 
Initially we have been using the axiom of \cite{SMIR} for 
our bound states, which in \cite{SMIR} had been established 
for first order poles only. However, it turns out that that 
this axiom can be applied to higher order poles as well 
provided one replaces the term ``res'' in (\ref{K-bound}) 
by simply taking the coefficient in the leading order 
singularity at the position of the pole. 

It is moreover remarkable that our result in the case of 
higher order poles is nonperturbative. It does not at 
all make use of the seperate Feynman diagrams which 
contribute to a singularity of the S-matrix \cite{BCDS}. This 
fact reflects the bootstrap property of the S-matrix 
on the level of form factors \cite{MAP}, see also 
\cite{DM2}. 

Since our calculation by (\ref{m-ident}) and (\ref{lamdef}) 
basically made use only of the fact that the S-matrix has the bootstrap 
property (\ref{S-bootstrap}), we conjecture that in 
other diagonal scattering theories with bootstrap property it should 
also be possible to reduce the problem of calculating 
form factors to one of evaluating recursive equations 
for polynomials.

\bigskip

{\bf 6.}

\bigskip

We are now going to determine the structure of the polynomials 
$Q_{[N_1,\ldots , N_r ]}({\bf{x}}^{(1)} , \ldots , {\bf{x}}^{(r)} )$. 

The first property is its total degree. This quantity is independent 
of the equations (\ref{kinemat}) and (\ref{bound}) and can be 
calculated by the following consideration.  

A Lorentz transformation shifts the rapidities simply by a parameter. 
We can therefore determine the effect of a Lorentz transformation 
on the denominator in (\ref{K-def}). The full form factor as defined 
in (\ref{formdef}) is required to be Lorentz invariant. Hence, 
the object $Q_{[N_1,\ldots , N_r ]}({\bf{x}}^{(1)}, \ldots , 
{\bf{x}}^{(r)} )$ must have the same behaviour under a 
Lorentz transformation as the denominator of $K$. A 
direct calculation using Proposition 1 and (\ref{m-ident}) leads 
to: 

\bigskip

{\bf Proposition 2:} {\it Let $M_{{\rm{sc}}}$ and $M_{{\rm{nsc}}}$ the 
set of all self-conjugate and of all non-self-conjugate particles 
respectively in an ADE Toda field theory. The total degree of the polynomial 
 $Q_{[N_1,\ldots , N_r ]}({\bf{x}}^{(1)}, \ldots , {\bf{x}}^{(r)} )$ 
is given by

\beq
\beqcol
{\rm deg}Q_{[N_1,\ldots , N_r ]}({\bf{x}}^{(1)}, \ldots , {\bf{x}}^{(r)} ) 
 &=& 
\sum\limits_{k=1}^{r} N_k(N_k-1) ( C^{-1}_{kk}+ C^{-1}_{k\bar{k}} - 
\delta_{k\bar{k}} ) \\
& & {\hspace{-4cm}} 
+ 2 \sum\limits_{k=1}^{r-1} \sum\limits_{l=k+1}^{r} 
  N_k N_l ( C^{-1}_{kl}+ C^{-1}_{k\bar{l}} - \delta_{k\bar{l}} ) 
+ {1\over{2}} \sum\limits_{k \in M_{{\rm{sc}}} } N_k ( N_k -1) 
+ {1\over{2}} \sum\limits_{k \in M_{{\rm{nsc}}} } N_k  N_{\bar{k}}.
\eeqcol
\label{prop2}
\eeq
 
}

\medskip

Since the coefficients in the recursion equations (\ref{kinemat}) 
and (\ref{bound}) are invariant under the transformation 
$ B-1 \to 1-B$ we find by induction in the numbers of particles:

\medskip

{\bf Proposition 3:} {\it The polynomial $
Q_{[N_1,\ldots , N_r ]}({\bf{x}}^{(1)}, \ldots , {\bf{x}}^{(r)} )$ 
is invariant under the weak-strong coupling duality 
$ B-1 \to 1-B$ . }

\bigskip

Let us introduce the elementary symmetric polynomials $e^n_r$ in 
$n$ variables $x_1, x_2, \ldots , x_n$ by \cite{MAC}

\beq
\prod\limits_{i=1}^n ( 1 + t x_i ) = \sum\limits_{l=0}^n 
  e^n_l t^l  .
\label{el-symm}
\eeq

Comparing this expression with the bracket symbol $[ p ] $ 
introduced in (\ref{brack-def-1}) we recognize that the 
recursion coefficients in (\ref{kinemat}) and (\ref{bound}) 
entirely consist of elementary symmetric polynomials. 

Given a partition $\pi = ( \pi_1, \pi_2, \ldots , \pi_p )$, 
with $\pi_1 \ge \pi_2 \ge \ldots \ge \pi_p$, 
we can define an elementary symmetric polynomial corresponding 
to this partition by

\beq
E^n_{\pi} = e^n_{\pi_1} e^n_{\pi_2} \cdots  e^n_{\pi_p} .
\label{el-symm-2}
\eeq

Characteristic quantities associated to a partition are its 
length $l(\pi )$, which is the number of nonzero entries 
of $\pi$, and its weight $ | \pi | = \sum_{i=1}^{p} \pi_i $. 

Since we are dealing with polynomials in $r$ different types 
of variables (\ref{vecdef}) we introduce the following polynomial 
which is symmetric only in each species of variables. 

\beq
E_{\Pi} ( {\bf{x}}^{(1)}, \ldots , {\bf{x}}^{(r)} ) = 
E_{ ( \pi^{(1)} | \pi^{(2)} | \ldots | \pi^{(r)} ) }
( {\bf{x}}^{(1)}, \ldots , {\bf{x}}^{(r)} ) = 
\prod\limits_{k=1}^{r} E_{\pi^{(k)}} ({\bf{x}}^{(k)}) .
\label{el-symm-3}
\eeq

Having in mind that the recursion coefficients in (\ref{kinemat}) 
and (\ref{bound}) consist just of polynomials of this kind, we find 
by induction in the number of particles:

\medskip

{\bf Proposition 4:} {\it 
\beq
Q_{[N_1, \ldots , N_r ]}  ( {\bf{x}}^{(1)}, \ldots , {\bf{x}}^{(r)} )
 = \sum\limits_{ | \Pi | = {\rm{deg}Q}} c_{\Pi} 
E_{\Pi} ( {\bf{x}}^{(1)}, \ldots , {\bf{x}}^{(r)} ), 
\label{prop4}
\eeq
where $c_{\Pi} $ are constants to be determined by the 
recursive equations (\ref{kinemat}) and (\ref{bound}).
}

\bigskip

We are now turning to the question which partitions out of the 
many possible ones in (\ref{prop4}) do actually contribute with 
a nonvanishing $c_{\Pi}$.
 
The quantity we would like to compute is the partial 
degree of the 
$Q_{[N_1, \ldots , N_r ]}  ( {\bf{x}}^{(1)}, \ldots , {\bf{x}}^{(r)} )$. 
This is the highest power to which any $x_i^{(k)}$ with 
$ i \in \{ 1,2, \ldots , N_k \} $ appears in the polynomial 
in question. Using Proposition 4 and (\ref{el-symm-2}), 
(\ref{el-symm-3}) it is clear that this partial degree 
of the variables of species $k$ is nothing else than 
the longest partition $\pi^{(k)}$ in $\Pi$. This means 
that for a given $N$-particle polynomial 
$Q_{[N_1, \ldots , N_r ]}  ( {\bf{x}}^{(1)}, \ldots , {\bf{x}}^{(r)} )$ 
we have to find the upper bounds for $l(\pi^{(k)} )$ for 
all $k \in I $. 

This upper bound can be calculated by investigating how the 
lengths of the partitions evolve when we add one variable 
to the polynomial $Q$, i.e. if we consider the recursion 
equation (\ref{bound}) which links an $N+2$-particle 
polynomial to an $N+1$-particle polynomial. In particular 
we need to know, according to the remark after 
equation (\ref{el-symm}), the content of partitions in 
the recursion coefficient in (\ref{bound}). 

\medskip

In order to do this we first determine the number $\nu^{c}_{ab;d}$ 
of $\langle p \rangle^{-1}$ entries in $\lambda^{c}_{ab;d} (\theta )$, 
introduced in (\ref{lamdef}). We take (\ref{lam-ident}) 
and count the number of $\langle p \rangle^{-1}$ entries on 
both sides of the equation. Using Proposition 1 we find 

\beq
 \nu^{b}_{ca;{\bar{d}}} + \nu^{c}_{ab;d} = (\lambda_a, \lambda_d) 
+  (\lambda_a, \lambda_{\bar{d}}) . 
\label{nu}
\eeq        
 
Considering all possible permutations of $a$, $b$, and $c$ 
and the exchange of $d$ and ${\bar{d}}$ in this equation we find 
that 

\beq
  \nu^{c}_{ab;d} = \left( \lambda_a +  \lambda_b - \lambda_{\bar{c}}, 
\lambda_d \right) .
\label{vu-2}
\eeq

The main entity of the recursion coefficient in (\ref{bound}) 
is the bracket symbol $ \{ p \}_i^{(k)}$ introduced in 
(\ref{brack-def-2}). The structure of this symbol is quite 
close to the one of $\lambda^{c}_{ab;d} (\theta )$. 
Taking the product $\prod_{i=1}^{N_k} \{ p \}^{(k)}_i $ 
we see using (\ref{brack-def-1}) and (\ref{el-symm}) that it can be expanded 
into elementary symmetric polynomials. The maximum 
length of the partitions in the $k$ component in the 
recursion coefficient in (\ref{bound}) can the be determined 
to be 

\beq
 2 \left( \lambda_a  + \lambda_b  - \lambda_{\bar{c}}  +  
          \lambda_{\bar{a}} +   \lambda_{\bar{b}}  - \lambda_c \; 
      , \; \lambda_k   \right) 
 + \delta_{ck} - \delta_{\bar{b}k} - \delta_{\bar{a}k} .
\label{rapf}
\eeq

This means that from one step in the recursion procedure, 
$l(\pi^{(k)})$ in the $N+2$-particle polynomial can be 
at most the maximal length of the partitions in the 
$N+1$-particle polynomial plus the expression in 
(\ref{rapf}).

Suppose we are given a particle vector ${\bf N} = [N_1, \ldots , 
N_r ]$ which by a series of successive fusings, with ${\bf N}_i$ 
being the intermediate vectors, terminates in a vector 
${\bf N}^{\prime}$. Hence we have 
${\bf N}^{\prime} < {\bf N}_1 < {\bf N}_2 < \ldots < {\bf N}_n < 
{\bf N}$, where the ordering is given by the absolute number 
of particles in a specific vector. 

We know that a particle in a Toda field theory corresponds 
to a fundamental weight vector. Moreover, we have seen in 
(\ref{vu-2}) that a fusing $a + b \to {\bar{c}}$ corresponds 
to a weight vector (conveniently to be taken in Dynkin 
components) $\lambda_a + \lambda_b - \lambda_{\bar{c}}$. 

Using this interpretation it is clear that the particle 
vectors introduced above are nothing else than vectors in 
the positive weight lattice of the Lie algebra ${\bf{g}}$. 
Any difference ${\bf{N}}_i - {\bf{N}}_j $ is in the 
positive weight lattice as long as $i > j$. 

Let us denote $\Lambda = {\bf{N}} - {\bf{N}}^{\prime}$. Rewriting 
the $\delta$'s in (\ref{rapf}) by expressions of the form 
$(\lambda_c , \alpha_k)  $ we get the following:

{\bf Lemma 1:} {\it Suppose we know, in the above notation, 
that the polynomial $Q_{{\bf{N}}}$ can be obtained by a polynomial 
$Q_{{\bf{N}}^{\prime}}$ by a process of successive fusings. Then the 
maximum length of the partitions belonging to the variables of type 
$k$ for any $k \in I $ are related by }

\beq
\beqcol
l(\pi^{(k)}) |_{{\bf{N}}} & = &
l(\pi^{(k)}) |_{{\bf{N}}^{\prime}} 
+ \sum_{i} \left( 2 (\lambda_{a_i} + \lambda_{b_i} - 
                     \lambda_{\bar{c}_i}, \lambda_k) 
+ 2 ( \lambda_{{\bar{a}}_i} + \lambda_{\bar{b}_i} - 
                     \lambda_{c_i}, \lambda_k )
-  (\lambda_{\bar{a}_i} + \lambda_{\bar{b}_i} - \lambda_{c_i} ) 
\right)                                                          \\
& &                                                               \\
 & = & l(\pi^{(k)}) |_{{\bf{N}}^{\prime}} 
 + 2 (\Lambda + \bar{\Lambda} , \lambda_k ) - 
      ( \bar{\Lambda} , \alpha_k ) .
\eeqcol
\label{gewicht}
\eeq

\bigskip

This result is interesting because it gives a link between 
the representations of the symmetric group acting on 
the polynomials $Q_{[N_1, \ldots , N_r]}$ and the 
weight space of the Lie algebra ${\bf{g}}$. Namely, the 
quantity $l(\pi^{(k)})$ restricts the representations 
of the symmetric group which do appear in the expression 
(\ref{prop4}).

\medskip 

We would like to rewrite the result (\ref{gewicht}) 
in more direct terms using the correspondence 
${\bf{N}} = \sum_{i=1}^{r} \lambda_i N_i $. 
By formula (\ref{prop2}) we 
know that the degree of $Q$-polynomial is zero 
if exactly one particle is present. Combining 
(\ref{gewicht}) and (\ref{prop2}) keeping in mind that 
we exclude to talk about a polynomial without any 
variables in our context we get: 

\medskip

{\bf Corollary:} {\it The maximum length of the partitions 
belonging to particles of type $k$ for any $k \in I $ 
in the polynomial $Q_{[N_1, \ldots , N_r ]}$ 
is given by:

\beq
l(\pi^{(k)} ) |_{[N_1,N_2, \ldots , N_r]} = 
 2  \sum_{i=1}^{r} N_i ( C^{-1}_{ik} + C^{-1}_{\bar{i} k} ) 
 - N_k \delta_{\bar{k} k} - 2 (C^{-1}_{kk} + C^{-1}_{\bar{k}k} - 
{1 \over {2}} \delta_{\bar{k} k} ).
\label{coroll}
\eeq
} 

We would like to close this paper with a conjecture on 
the minimal length of the partitions contributing 
nonvanishingly to (\ref{prop4}). We rewrite 
(\ref{K-def}) as $K_{[N_1, \ldots , N_r]} = 
Q_{[N_1, \ldots , N_r]} / D_{[N_1, \ldots , N_r]} $. 
By construction we have as in (\ref{prop4}) 

\beq
 D_{[N_1, \ldots , N_r]} ( {\bf{x}}^{(1)}, \ldots , {\bf{x}}^{(r)} )
 = \sum\limits_{ | \Xi | = {\rm{deg}Q_{[N_1, \ldots , N_r]}}} d_{\Xi} 
E_{\Xi} ( {\bf{x}}^{(1)}, \ldots , {\bf{x}}^{(r)} ), 
\label{D-def}
\eeq

where all constants $d_{\Xi}$ are of course fixed by the 
explicit form of $ D_{[N_1, \ldots , N_r]} $ in (\ref{K-def}). 

By construction it is clear that the polynomial 
$ D_{[N_1, \ldots , N_r]} $ is just the kernel of 
the recursive equations (\ref{kinemat}) and (\ref{bound}) 
and has therefore to be added to any particular solution 
to these equations in order to have the most general 
solution for any $ Q_{[N_1, \ldots , N_r]} $. This, of course, 
introduces a free parameter at each stage of the recursion 
process. 

Let $l_{{\rm{min}}}(\xi^{(k)})$ be the minimum length of 
of the partitions $\xi^{(k)}$, belonging to variables 
of species $k$, for any $ k \in I $, 
contributing nonvanishingly to the sum (\ref{D-def}). 

Based on calculating solutions to the equations 
(\ref{kinemat}) and (\ref{bound}) we would like to 
state the following: 

{\it Conjecture:} The partitions $\pi^{(k)}$ contributing 
nonvanishingly to the a general solution 
$Q_{[N_1, \ldots , N_r]}$ in the form of (\ref{prop4}) are 
at least equally long as $l_{{\rm{min}}}(\xi^{(k)})$, 
for any $ k \in I $. We would like to 
put it even stronger. The partitions 
$\Pi $ in (\ref{prop4}) do contribute nonvanishingly to 
a general solution $Q_{[N_1, \ldots , N_r]}$ if and only if the 
length of its subpartitions $\pi^{(k)}$ 
is between this lower bound and the upper bound 
(\ref{coroll}) for $k\in I $.

\medskip

It is possible to compute $l_{{\rm{min}}}(\xi^{(k)})$ explicitly:

\beq
l_{{\rm{min}}} (\xi^{(k)} ) \ge {1\over {2}} (N_k - 1 ) \rfloor 
 + \delta_{k\bar{k}} ( C^{-1}_{kk} + C^{-1}_{k \bar{k}} -1 ) 
   (N_k -1 ),
\eeq

where $1/2 (N_k - 1 ) \rfloor $ means that if this expression is 
not an integer we have to take the next integer. 

\bigskip

{\bf Acknowledgement:} The author is grateful to R. Sasaki 
and to A. Pressley for helpful discussions.

\end{document}